\documentclass[prd,preprint,showpacs,superscriptaddress,nofootinbib,floatfix]{revtex4}
\usepackage{latexsym}
\usepackage{epsfig}

\begin{document}

\preprint{ILL-(TH)-02-12}
\preprint{RM3-TH/02-21}
\preprint{FERMILAB-Pub-02/341-T}

\title{Higgs-Boson Production via Bottom-Quark Fusion}

\author{F.~Maltoni}
\affiliation{Centro Studi e Ricerche ``Enrico Fermi''\break
via Panisperna 89/A, 00184 Rome, Italy}
\altaffiliation[Mail address: ]{Dipartimento di Fisica, Terza Universit\`a
di Roma, via della Vasca Navale 84, 00146 Rome, Italy}
\author{Z.~Sullivan}
\affiliation{Theoretical Physics Department, Fermi National
Accelerator Laboratory,\hspace*{-1in}\\ Batavia, IL, 60510-0500}
\author{S.~Willenbrock}
\affiliation{Theoretical Physics Department, Fermi National Accelerator
Laboratory,\hspace*{-1in}\\ Batavia, IL, 60510-0500}
\affiliation{Department
of Physics, University of Illinois at Urbana-Champaign, 1110 West Green
Street, Urbana, IL 61801}

\date{January 6, 2003}

\begin{abstract}
Higgs bosons with enhanced coupling to bottom quarks are copiously produced at
hadron colliders via $b\bar b\to h$, where the initial $b$ quarks reside in
the proton sea.  We revisit the calculation of the next-to-leading-order cross
section for this process and argue that the appropriate factorization scale
for the $b$ distribution functions is approximately $m_h/4$, rather than $m_h$,
as had been previously assumed.  This greatly improves the convergence of the
perturbation series, and yields a result with mild factorization-scale
dependence. We also show that the leading-order calculation of $gg\to b\bar
bh$, integrated over the momenta of the final-state particles, is very
sensitive to the factorization and renormalization scales.  For scales of
order $m_h/4$ the $gg\to b\bar bh$ cross section is comparable to that of
$b\bar b\to h$, in contrast to the order-of-magnitude discrepancy between
these two calculations for the scale $m_h$.  The result we obtain improves the
prospects for Higgs-boson discovery at hadron colliders for large values of
$\tan\beta$.
\end{abstract}


\maketitle

\section{Introduction}
\label{sec:intro}

In the standard model, the Higgs boson couples to fermions with strength
$m_f/v$, where $v=(\sqrt 2 G_F)^{-1/2} \approx 246$ GeV is the
vacuum-expectation value of the Higgs field. The Yukawa coupling of the Higgs
boson to bottom quarks ($m_b \approx 5$ GeV) is thus very weak, leading to very
small cross sections for associated production of the Higgs boson and bottom
quarks at the Fermilab Tevatron ($\sqrt S=1.96$ TeV $p\bar p$)
\cite{Stange:ya} and the CERN Large Hadron Collider (LHC, $\sqrt S=14$ TeV
$pp$) \cite{Dicus:1988cx}. However, this Yukawa coupling could be considerably
enhanced in extensions of the standard model with more than one Higgs doublet,
thereby increasing this production cross section \cite{Dicus:1988cx}.  For
example, in a two-Higgs-doublet model, the Yukawa coupling of some or all of
the Higgs bosons ($h^0,H^0,A^0,H^\pm$) to the bottom quark could be enhanced
for large values of $\tan\beta = v_2/v_1$, where $v_1$ is the vacuum
expectation value of the Higgs doublet that couples to the bottom quark.

\begin{figure}[t]
\begin{center}
\vspace*{0cm} \hspace*{0cm} \epsfxsize=4cm \epsfbox{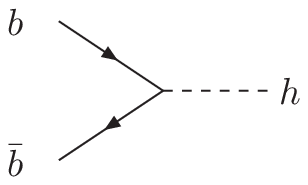} \vspace*{-.8cm}
\end{center}
\caption{Leading-order diagram for the production of the Higgs boson via
bottom-quark fusion.\label{fig:bbh}}
\end{figure}

The dominant subprocess for the production of a Higgs boson in association
with bottom quarks is bottom-quark fusion, $b\bar b\to h$
(Fig.~\ref{fig:bbh}),\footnote{We use $h$ to denote a generic Higgs boson. In
a two-Higgs-doublet model, $h$ may denote any of the neutral Higgs bosons
($h^0,H^0,A^0$).} where the $b$ quarks reside in the proton sea
\cite{Dicus:1988cx,Dicus:1998hs,Balazs:1998sb}.  This is the leading-order
(LO) subprocess for the inclusive production of the Higgs boson in association
with bottom quarks.  Since the bottom-quark sea is generated by gluons
splitting into nearly-collinear $b\bar b$ pairs, the final state contains two
spectator bottom quarks that tend to be at low transverse momentum ($p_T$).

In contrast, if one requires one bottom quark at high $p_T$ from the
production process, the leading-order subprocess is $bg\to bh$
\cite{Choudhury:1998kr,Huang:1998vu,Campbell:2002zm}. This process is
particularly promising due to the ability to tag the $b$ quark in the final
state.  The cross section for the production of the Higgs boson accompanied by
two high-$p_T$ $b$ quarks is obtained at LO from the subprocesses $gg,q\bar
q\to b\bar bh$
\cite{Dai:1994vu,Dai:1996rn,Richter-Was:1997gi,Drees:1997sh,Diaz-Cruz:1998qc,
Balazs:1998nt,Carena:1998gk,Dawson:2002cs}.\footnote{$q\bar q\to b\bar bh$ is
negligible in comparison with $gg\to b\bar bh$.}  Although this process has
been the most studied, it is likely that $bg\to bh$ is the more promising, due
to its larger cross section.  The inclusive cross section, $b\bar b\to h$,
which we study in this paper, is useful when the Higgs boson can be identified
above backgrounds without the need to detect the accompanying bottom quarks
that reside in the final state. This subprocess may be useful to discover a
Higgs boson for large $\tan\beta$ in the decay mode $h\to \tau^+\tau^-$ at the
Tevatron and LHC \cite{Kunszt:1991qe,unknown:1999fr} and $h\to \mu^+\mu^-$ at
the LHC \cite{unknown:1999fr,Kao:1995gx,Barger:1997pp}. It has the advantage
of having the largest cross section, since it is inclusive of the other two
processes.

\begin{figure}[t]
\begin{center}
\vspace*{0cm} \hspace*{0cm} \epsfxsize=10cm \epsfbox{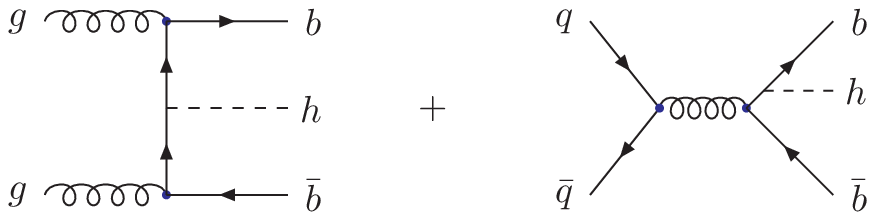} \vspace*{-.8cm}
\end{center}
\caption{Representative diagrams for associated production of the Higgs boson
and two high-$p_T$ bottom quarks: (a) $gg\to b\bar bh$ (8~diagrams); (b) $q\bar
q\to b\bar bh$ (2~diagrams).\label{fig:ggbbh}}
\end{figure}

Higgs-boson production via bottom-quark fusion was calculated at
next-to-leading order (NLO) in Refs.~\cite{Dicus:1998hs,Balazs:1998sb}.  There
are two puzzling aspects of the results of that calculation:
\begin{itemize}

\item Although the NLO correction is modest, it consists of two independent
corrections, of order $1/\ln(m_h/m_b)$ and $\alpha_S$, which are both large
(and of opposite sign).  This suggests that the perturbation series in each
expansion parameter individually may not be well behaved.

\item The cross section at the Tevatron (both LO and NLO) is an order
of magnitude larger than the cross section obtained by calculating $gg\to
b\bar bh$ (Fig.~\ref{fig:ggbbh}) and integrating over the momenta of the
final-state particles \cite{Carena:2000yx}. While $gg\to b\bar bh$ is not a
reliable calculation of the total inclusive cross section, since the expansion
parameter is $\alpha_S\ln(m_h/m_b)$ rather than $\alpha_S$
\cite{Dicus:1998hs}, the large discrepancy between the two calculations is
surprising.

\end{itemize}

In this paper we solve both of these puzzles.  Implicit in both puzzles is the
choice of the factorization scale, which had been chosen to be the Higgs-boson
mass in Refs.~\cite{Dicus:1998hs,Balazs:1998sb}.  Although the choice of the
factorization scale in a fixed-order calculation is often regarded as
arbitrary, we argue that there is a prescription based on physical
considerations.  Refining the discussion of Ref.~\cite{Plehn:2002vy}, we show
that the relevant factorization scale is a fraction of the Higgs-boson mass,
approximately $m_h/4$. We find that this choice of scale solves both puzzles
listed above. We thereby present a reliable NLO calculation of the inclusive
cross section for the production of the Higgs boson in association with bottom
quarks.

In Section~\ref{sec:scale} we determine the relevant factorization scale for
Higgs-boson production in association with bottom quarks.  In
Section~\ref{sec:results} we present the results of this scale choice.  We
discuss these results in Section~\ref{sec:discussion}, and show that they
solve the two puzzles listed above.  Section~\ref{sec:conclusions} summarizes
the conclusions of our study and suggests further work.

\section{Factorization scale}\label{sec:scale}

The $b$ distribution function, like any parton distribution function, sums (to
all orders) collinear logarithms that appear at higher orders.  Thus, to
determine the relevant factorization scale, we investigate the collinear
logarithm that arises at next-to-leading order.

\begin{figure}[b]
\begin{center}
\vspace*{0cm} \hspace*{0cm} \epsfxsize=10cm \epsfbox{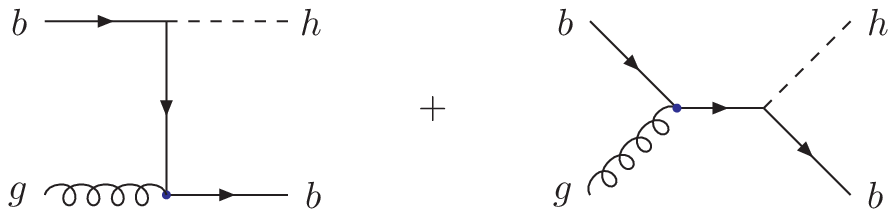} \vspace*{-.8cm}
\end{center}
\caption{Diagrams for the next-to-leading-order correction to $b\bar b\to h$
from initial gluons.  This correction is of order $1/\ln(m_h/m_b)$.
\label{fig:gbhb}}
\end{figure}

\begin{figure}[t]
\begin{center}
\vspace*{0cm} \hspace*{0cm} \epsfxsize=12cm \epsfbox{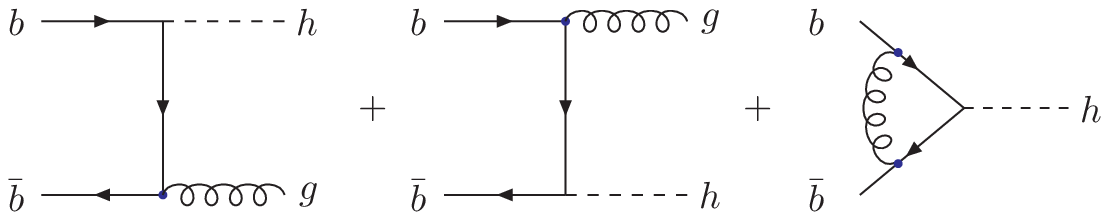} \vspace*{-.8cm}
\end{center}
\caption{Diagrams for the next-to-leading-order correction to $b\bar b\to h$
from real and virtual gluon emission.  This correction is of order
$\alpha_S$.\label{fig:bbhvirtual}}
\end{figure}

There are two independent NLO corrections to $b\bar b\to h$.  The first is from
initial gluons, $bg\to bh$ (Fig.~\ref{fig:gbhb}), which is a correction of
order $1/\ln(m_h/m_b)$, as explained in Ref.~\cite{Dicus:1998hs}.  The second
is from virtual and real gluon emission (Fig.~\ref{fig:bbhvirtual}), which is a
correction of order $\alpha_S$.  Since $\alpha_S(m_h)$ is proportional to
$1/\ln(m_h/\Lambda_{QCD})$, one can regard both of these corrections as being
of the form of an inverse logarithm.

The calculations involved are identical to those of
Refs.~\cite{Dicus:1998hs,Balazs:1998sb}. We keep the effect of the bottom-quark
mass exactly in our calculations, without approximations. In order to do this,
we may set the bottom-quark mass to zero in all diagrams in which the bottom
quark appears as an initial-state parton.\footnote{One may maintain a nonzero
bottom-quark mass in these diagrams, but it does not increase the accuracy of
the calculation.} This is called the simplified Aivazis-Collins-Olness-Tung
(ACOT) scheme \cite{Aivazis:1993pi,Collins:1998rz,Kramer:2000hn}. The only
subprocess in which we must keep the bottom-quark mass is $gg\to b\bar bh$
(Fig.~\ref{fig:ggbbh}), which is a next-to-next-to-leading-order correction of
order $1/\ln^2(m_h/m_b)$.  In practice, for $m_h\gg m_b$, it is an excellent
approximation to neglect the bottom-quark mass in this subprocess; we keep the
mass nonzero, nevertheless.

Let us first investigate the $1/\ln(m_h/m_b)$ correction, from initial gluons.
The first diagram in Fig.~\ref{fig:gbhb} has a collinear divergence due to the
$t$-channel quark propagator.  The hadronic differential cross section
therefore has the behavior $d\sigma/dt\sim 1/t$ in the collinear region.  The
integral over $t$ produces the collinear logarithm.  Thus, the upper limit of
the collinear integration is set by the virtuality, $\sqrt{-t}$, at which the
differential cross section begins to deviate substantially from the collinear
behavior.

\begin{figure}[t]
\begin{center}
\vspace*{0cm} \hspace*{0cm} \epsfxsize=10cm \epsfbox{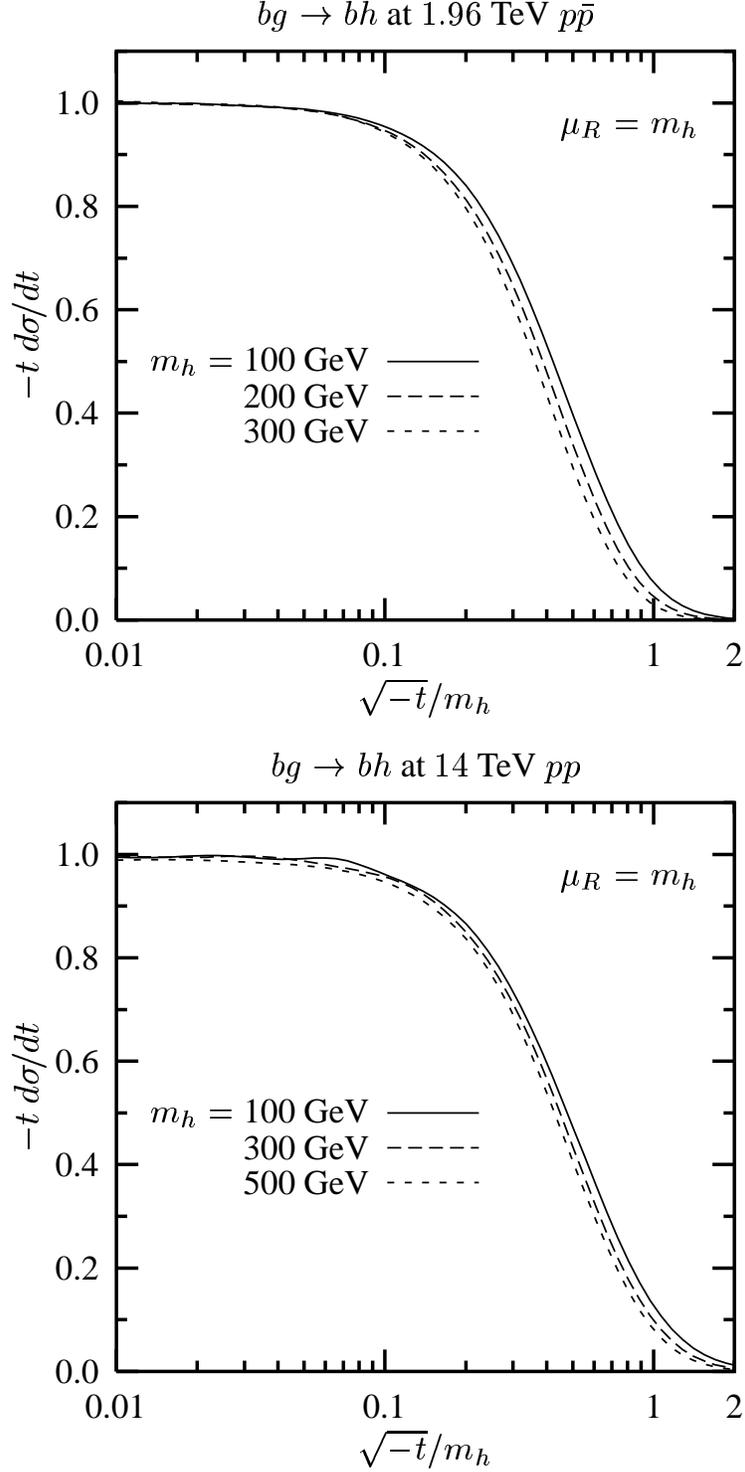}
\vspace*{-.8cm}
\end{center}
\caption{Hadronic differential cross section times the squared virtuality for
the subprocess $bg\to bh$ {\it vs.}\ the virtuality (scaled to the Higgs-boson
mass) at both the Tevatron (upper plot) and the LHC (lower plot). Curves are
shown for a variety of Higgs-boson masses, scaled such that they overlap at
small virtuality.\label{fig:dsigdtgbhb}}
\end{figure}

We show in Fig.~\ref{fig:dsigdtgbhb} the hadronic differential cross section
times the squared virtuality, $-t\,d\sigma/dt$, versus the virtuality (scaled
to the Higgs-boson mass), $\sqrt{-t}/m_h$, for a variety of Higgs-boson masses
at the Tevatron and the LHC. In order to compare the cross sections for
different Higgs-boson masses at a given collider, we normalize the curves to
unity at small virtualities. The curves are nearly identical, demonstrating
that the differential cross section scales with the Higgs-boson mass.  At
small virtualities the curve is flat, indicating the collinear behavior
$d\sigma/dt\sim 1/t$. At larger virtualities the cross section is damped.  For
fixed $t$, the differential cross section is given by \cite{Campbell:2002zm}
\begin{equation}
\frac{d\sigma}{dt}=-\frac{1}{S}\int^S_{m_h^2-t}ds\int_{{1\over 2}\ln
(s/S)}^{-{1\over 2}\ln (s/S)} d\eta\,
[g(x_1,\mu_F)b(x_2,\mu_F)+(x_1\leftrightarrow x_2)]
\frac{\alpha_S(\mu_R)}{24}\left(\frac{y_b(\mu_R)}{\sqrt
2}\right)^2\frac{1}{s^2}\frac{m_h^4+u^2}{st}
\end{equation}
which explicitly shows the $1/t$ behavior for small $t$.  For larger values of
$-t$, the lower limit on the $s$ integration, $m_h^2-t$, increases and damps
the cross section, since the integrand falls steeply with increasing $s$.

Figure~\ref{fig:dsigdtgbhb} shows that the virtuality at which the behavior of
the differential cross section deviates substantially from the collinear
behavior is much less than $m_h$. Since the curves vary smoothly, there is
some ambiguity in defining the virtuality at which the collinear behavior
ceases. For the sake of discussion, let us define $\sqrt{-t}\leq m_h/4$ as the
collinear region. The collinear region extends to slightly higher virtualities
at the LHC than at the Tevatron, but this is small compared with the inherent
ambiguity in defining the collinear region.

The collinear logarithm that is generated at NLO is therefore approximately
$\ln(m_h/4\mu_F)$, rather than $\ln(m_h/\mu_F)$. Thus the factorization scale
for the $1/\ln(m_h/m_b)$ correction should be chosen to be of order $\mu_F
\approx m_h/4$ in order to sum the collinear logarithm.  We will examine the
consequences of this scale choice in the next section.  In order to account
for the ambiguity in defining the collinear region, we will vary the
factorization scale between $\mu_F=m_h/8$ and $\mu_F=m_h/2$.

This derivation of the factorization scale is similar to the argument
presented in Ref.~\cite{Plehn:2002vy}, where the behavior of the cross section
as a function of the transverse momentum of the final-state $b$ quark,
$p_{T,b}$, is studied.  We prefer to instead use the variable $\sqrt{-t}$,
which has the interpretation of the virtuality of the $t$-channel bottom-quark
propagator. Since we adopt the simplified ACOT formalism
\cite{Aivazis:1993pi,Collins:1998rz,Kramer:2000hn}, we are able to set the $b$
mass to zero, which makes the discussion of the finite $b$ mass in
Ref.~\cite{Plehn:2002vy} moot. Despite these differences, our approach is very
similar to that of Ref.~\cite{Plehn:2002vy} and yields similar results.

\begin{figure}[t]
\begin{center}
\vspace*{0cm} \hspace*{0cm} \epsfxsize=10cm \epsfbox{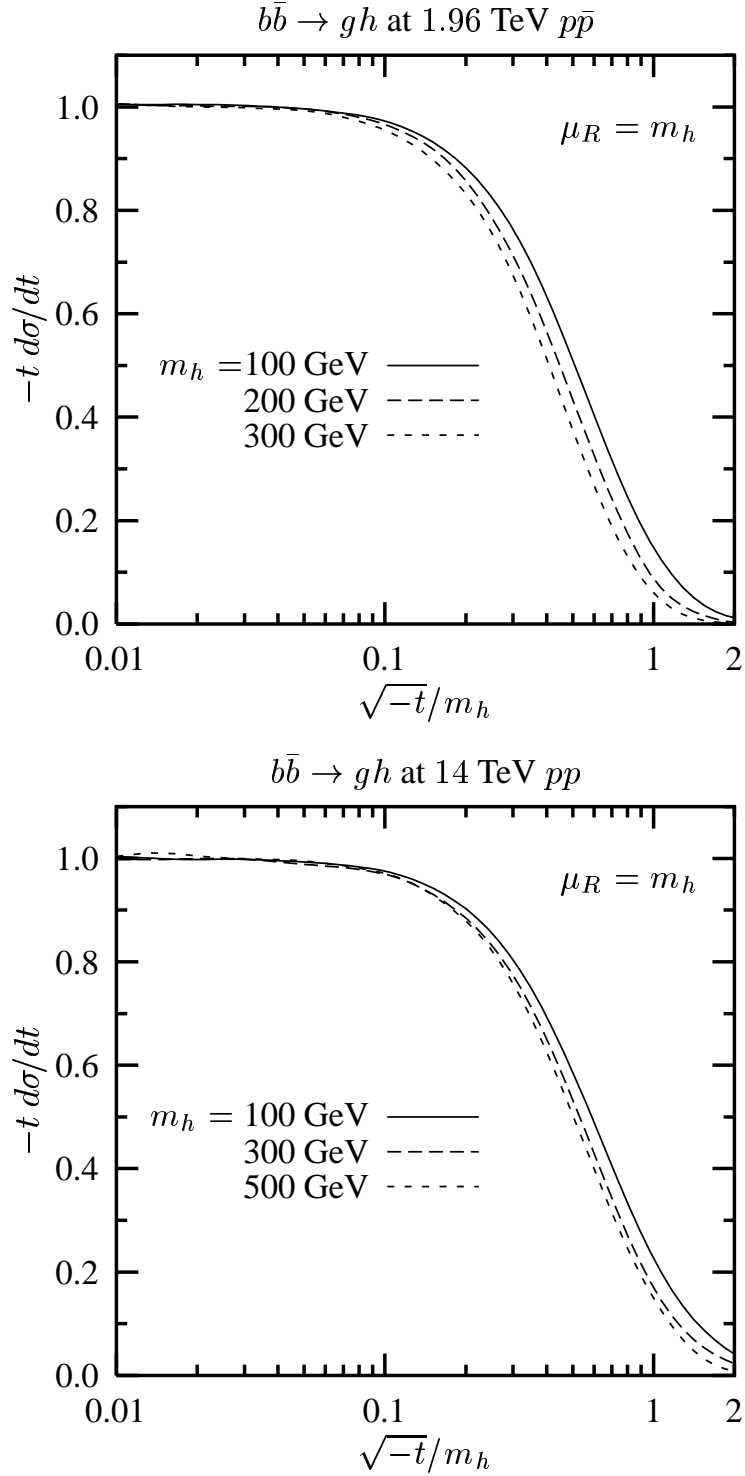}
\vspace*{-.8cm}
\end{center}
\caption{Same as Fig.~\ref{fig:dsigdtgbhb}, but for the subprocess $b\bar b\to
gh$.  A cut of $-u > (10\;{\rm GeV})^2$ is imposed to regulate the infrared
singularity associated with soft-gluon emission.\label{fig:dsigdtbbhg}}
\end{figure}

The results for the $\alpha_S$ correction from virtual and real gluon emission
(Fig.~\ref{fig:bbhvirtual}) are similar.  We show in Fig.~\ref{fig:dsigdtbbhg}
the hadronic differential cross section for real gluon emission times the
squared virtuality, $-t\,d\sigma/dt$, versus the virtuality (scaled to the
Higgs-boson mass), $\sqrt{-t}/m_h$, for a variety of Higgs-boson masses at the
Tevatron and the LHC.  A cut of $-u>(10\;{\rm GeV})^2$ is imposed to regulate
the infrared singularity associated with soft-gluon emission.  The collinear
region extends up to a slightly higher virtualities than in the $bg\to bh$
subprocess, but the difference is small compared with the inherent ambiguity
in defining the collinear region.  The collinear region again extends to
slightly higher virtualities at the LHC than at the Tevatron.

\section{Results}\label{sec:results}

Following Refs.~\cite{Dicus:1998hs,Balazs:1998sb}, we fix the renormalization
scale of the Yukawa coupling to $\mu_R=m_h$.  It was shown in that paper that
the renormalization-scale dependence of the cross section is modest, and is
reduced at NLO in comparison with LO.  Hence we focus on the
factorization-scale dependence of the cross section.

\begin{figure}[htb]
\begin{center}
\vspace*{0cm} \hspace*{0cm} \epsfxsize=10cm \epsfbox{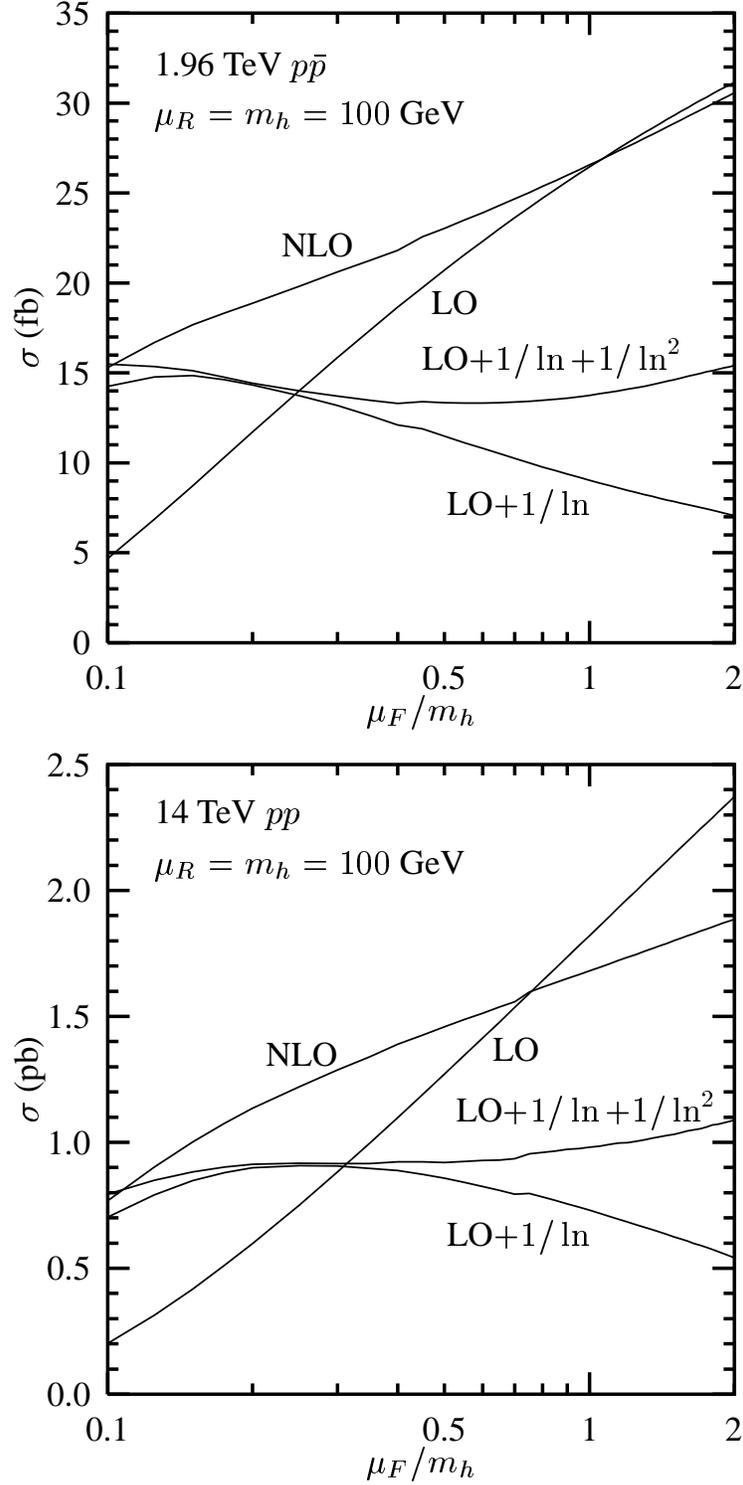}
\vspace*{-.8cm}
\end{center}
\caption{Cross section for Higgs-boson production via bottom-quark
fusion {\it vs.}\ the factorization scale for $m_h=100$ GeV at the
Tevatron (upper plot) and the LHC (lower plot).  Curves are shown for
leading-order, next-to-leading order, leading-order plus corrections
of order $1/\ln(m_h/m_b)$, and leading-order plus corrections of order
$1/\ln(m_h/m_b)$ and $1/\ln^2(m_h/m_b)$.\label{fig:sigma}}
\end{figure}

We show in Fig.~\ref{fig:sigma} the factorization-scale dependence of the
inclusive cross section for Higgs-boson production via bottom-quark fusion for
$m_h=100$ GeV at the Tevatron and the LHC. The four curves on each plot are
described below.  The results for heavier Higgs bosons are qualitatively
similar. While a standard-model Higgs boson of 100 GeV is excluded, the lower
bounds on the masses of the neutral Higgs bosons $h^0,A^0$ of the minimal
supersymmetric standard model are about 91 GeV \cite{:2001xx}. Even lighter
Higgs bosons are possible in a general two-Higgs-doublet model \cite{OPAL}. We
set $\tan\beta=1$ throughout.

Let us first focus on the results at the Tevatron, the upper plot in
Fig.~\ref{fig:sigma}. The factorization scale covers a wide range, including
the canonical choice $\mu_F= m_h$ used in
Refs.~\cite{Dicus:1998hs,Balazs:1998sb} and $\mu_F\approx m_h/4$ advocated in
the previous section. The curve labeled ``LO'' is the LO cross section
calculated with LO parton distribution functions (CTEQ6L1
\cite{Pumplin:2002vw}), which has significant factorization-scale dependence.
The curve labeled ``LO$+1/\ln$'' is the partial NLO cross section, including
only the $1/\ln(m_h/m_b)$ correction, calculated with NLO parton distribution
functions (CTEQ6M). At $\mu_F = m_h$ this correction is large and negative,
approximately $-70\%$. However, at $\mu_F\approx m_h/4$ this correction is
small, indicating that this is indeed the relevant factorization scale for
this process.

The curve labeled ``NLO'' in Fig.~\ref{fig:sigma} is the full NLO cross
section, including both the $1/\ln(m_h/m_b)$ correction and the $\alpha_S$
correction, calculated with NLO parton distribution functions. At $\mu_F =
m_h$, the $\alpha_S$ correction is large and positive, nearly canceling the
large negative $1/\ln(m_h/m_b)$ correction. However, at $\mu_F\approx m_h/4$,
the $\alpha_S$ correction is modest, yielding a modest NLO correction. This
again indicates that this is the relevant factorization scale for this
process.  The factorization-scale dependence of the NLO cross section is
reduced in comparison with that of the LO cross section.

The curve labeled ``LO$+1/\ln+1/\ln^2$'' in Fig.~\ref{fig:sigma} is the partial
NLO cross section [LO plus $1/\ln(m_h/m_b)$ correction] plus the
$1/\ln^2(m_h/m_b)$ correction from the diagrams in
Fig.~\ref{fig:ggbbh},\footnote{The correction from $q\bar q\to b\bar bh$ is
negligible in comparison with that from $gg\to b\bar bh$.} which is part of the
next-to-next-to-leading-order (NNLO) correction.\footnote{We calculated this
curve with NLO parton distribution functions since NNLO parton distribution
functions are not yet available.} This NNLO correction is smallest at
$\mu_F\approx m_h/4$, again indicating that this is the relevant factorization
scale for this process. This is significant because while it is always
possible to find a factorization scale such that the NLO correction is small,
it is not guaranteed that this same scale will yield a small NNLO correction,
unless there is a good motivation for this scale.  We anticipate that a full
NNLO calculation will further support our argument that the relevant
factorization scale for this process is $\mu_F\approx m_h/4$.

The results at the LHC, the lower plot in Fig.~\ref{fig:sigma}, are
qualitatively similar to those at the Tevatron.  We argued in the previous
section that the relevant factorization scale for this process at the LHC is
slightly higher than at the Tevatron.  The partial NLO cross section [LO plus
$1/\ln(m_h/m_b)$ correction] crosses the LO cross section at a slightly higher
factorization scale at the LHC than at the Tevatron, consistent with this
argument.

We present in Tables~\ref{tab:sigma1}--\ref{tab:sigma3} the NLO cross sections
for Higgs-boson production via bottom-quark fusion at the Tevatron (both
$\sqrt S=1.8$ and $1.96$ TeV) and the LHC, using $\mu_F=m_h/4$. These cross
sections differ from those of Refs.~\cite{Dicus:1998hs,Balazs:1998sb} in part
due to the improved choice of factorization scale, and also in part due to a
bug in the CTEQ4M \cite{Lai:1996mg} computer code that affected the gluon and
$b$ distribution functions used in that paper.  Since there is some ambiguity
in defining the collinear region, we vary the factorization scale between twice
and one-half its central value, and consider this an uncertainty in our
calculation.  This corresponds to the first uncertainty listed in
Tables~\ref{tab:sigma1}--\ref{tab:sigma3}. We also vary the renormalization
scale between $m_h/2$ and $2m_h$, and report this as the second uncertainty in
Tables~\ref{tab:sigma1}--\ref{tab:sigma3}.  This uncertainty is considerably
less than that associated with the factorization scale.

There are two additional sources of uncertainty in our calculation.  The
uncertainty in the $b$-quark $\overline{\rm MS}$ mass, $m_b(m_b)=4.24\pm 0.11$
\cite{El-Khadra:2002wp}, yields an uncertainty in the Yukawa coupling
(evaluated at $\mu_R=m_h$).  This corresponds to the third uncertainty in
Tables~\ref{tab:sigma1}--\ref{tab:sigma3}, which is the same for all
Higgs-boson masses and machine energies. The fourth uncertainty corresponds to
the uncertainty in the parton distribution functions, which we evaluated using
the method described in Refs.~\cite{Pumplin:2002vw,Sullivan:2002jt}.
The four sources of uncertainty are combined in quadrature and reported as an
absolute uncertainty on the NLO cross section.  The same exercise is performed
for the LO cross section, where we report only the combined uncertainty.

\begin{table}[tbp]
\begin{center}
\vspace*{-.6in}
\caption{Leading-order (LO) and next-to-leading-order (NLO) cross sections
(pb) for Higgs-boson production via bottom-quark fusion at the Tevatron ($\sqrt
S = 1.8$ TeV $p\bar p$).  The LO cross sections are computed using CTEQ6L1
parton distribution functions \cite{Pumplin:2002vw} and 1-loop evolution of the
bottom-quark Yukawa coupling, $y_b(\mu_R)$.  The NLO cross sections are
computed using CTEQ6M parton distribution functions and 2-loop evolution of the
Yukawa coupling and $\alpha_S(\mu_R)$.  The four sources of uncertainty in the
NLO cross sections are also listed.  The factorization scale is $\mu_F =
m_h/4$, and is varied between $\mu_F=m_h/8$ (upper uncertainty) and
$\mu_F=m_h/2$ (lower uncertainty).  The renormalization scale is $\mu_R=m_h$,
and is varied between $\mu_R=m_h/2$ (upper uncertainty) and $\mu_F=2m_h$
(lower uncertainty).  The uncertainty in the Yukawa coupling stems from the
uncertainty in the $b$ mass, $m_b=4.24\pm 0.11$ GeV. The final uncertainty is
due to the uncertainty in the parton distribution functions.  These four
uncertainties are combined in quadrature and reported as an absolute
uncertainty in the NLO cross section.  The combined uncertainty in the LO
cross section is also given. \label{tab:sigma1}}
\medskip
\begin{tabular}{cr@{ }l@{ }lr@{ }l@{ }lllll} \hline \hline
\multicolumn{1}{c}{$m_h$ (GeV)} & \multicolumn{3}{c}{$\sigma_{\rm LO}$ (pb)} &
\multicolumn{3}{c}{\quad$\sigma_{\rm NLO}$ (pb)} &
\multicolumn{4}{c}{$\delta\sigma_{\rm NLO}(\delta\mu_F, \delta\mu_R, \delta
y_b,$ PDF) (\%)} \\ \hline
  60 & 8.24 & $^{+7.44}_{-6.33}$ & $\times 10^{-2}$ & \quad
1.13 & $^{+0.33}_{-0.43}$ & $\times 10^{-1}$ & \quad\quad $^{-37}_{+28}$ &
$\pm  5.1$ & $\pm  6.4$ & $^{ +6.1}_{ -6.4}$ \\
  70 & 4.89 & $^{+3.63}_{-3.36}$ & $\times 10^{-2}$ & \quad
6.54 & $^{+1.67}_{-1.96}$ & $\times 10^{-2}$ & \quad\quad $^{-28}_{+23}$ &
$\pm  4.7$ & $\pm  6.4$ & $^{ +7.3}_{ -7.4}$ \\
  80 & 2.99 & $^{+1.89}_{-1.86}$ & $\times 10^{-2}$ & \quad
3.92 & $^{+0.91}_{-0.98}$ & $\times 10^{-2}$ & \quad\quad $^{-22}_{+20}$ &
$\pm  4.4$ & $\pm  6.4$ & $^{ +8.8}_{ -8.5}$ \\
  90 & 1.87 & $^{+1.05}_{-1.07}$ & $\times 10^{-2}$ & \quad
2.43 & $^{+0.53}_{-0.53}$ & $\times 10^{-2}$ & \quad\quad $^{-18}_{+18}$ &
$\pm  4.3$ & $\pm  6.4$ & $^{+10}_{-10}$ \\
 100 & 1.21 & $^{+0.61}_{-0.64}$ & $\times 10^{-2}$ & \quad
1.55 & $^{+0.33}_{-0.32}$ & $\times 10^{-2}$ & \quad\quad $^{-16}_{+16}$ &
$\pm  4.1$ & $\pm  6.4$ & $^{+12}_{-11}$ \\
 105 & 9.75 & $^{+4.69}_{-5.01}$ & $\times 10^{-3}$ & \quad
1.25 & $^{+0.27}_{-0.25}$ & $\times 10^{-2}$ & \quad\quad $^{-14}_{+15}$ &
$\pm  4.0$ & $\pm  6.4$ & $^{+13}_{-12}$ \\
 110 & 7.92 & $^{+3.67}_{-3.95}$ & $\times 10^{-3}$ & \quad
1.02 & $^{+0.22}_{-0.20}$ & $\times 10^{-2}$ & \quad\quad $^{-13}_{+14}$ &
$\pm  3.9$ & $\pm  6.4$ & $^{+14}_{-12}$ \\
 115 & 6.48 & $^{+2.90}_{-3.13}$ & $\times 10^{-3}$ & \quad
8.28 & $^{+1.81}_{-1.61}$ & $\times 10^{-3}$ & \quad\quad $^{-12}_{+14}$ &
$\pm  3.8$ & $\pm  6.4$ & $^{+15}_{-13}$ \\
 120 & 5.32 & $^{+2.31}_{-2.51}$ & $\times 10^{-3}$ & \quad
6.79 & $^{+1.51}_{-1.31}$ & $\times 10^{-3}$ & \quad\quad $^{-12}_{+13}$ &
$\pm  3.7$ & $\pm  6.4$ & $^{+16}_{-14}$ \\
 125 & 4.38 & $^{+1.85}_{-2.01}$ & $\times 10^{-3}$ & \quad
5.60 & $^{+1.27}_{-1.09}$ & $\times 10^{-3}$ & \quad\quad $^{-11}_{+13}$ &
$\pm  3.6$ & $\pm  6.4$ & $^{+17}_{-14}$ \\
 130 & 3.63 & $^{+1.50}_{-1.63}$ & $\times 10^{-3}$ & \quad
4.63 & $^{+1.08}_{-0.91}$ & $\times 10^{-3}$ & \quad\quad $^{-10}_{+12}$ &
$\pm  3.6$ & $\pm  6.4$ & $^{+18}_{-15}$ \\*
 140 & 2.52 & $^{+1.01}_{-1.09}$ & $\times 10^{-3}$ & \quad
3.22 & $^{+0.80}_{-0.65}$ & $\times 10^{-3}$ & \quad\quad $^{ -9.0}_{+12}$ &
$\pm  3.4$ & $\pm  6.4$ & $^{+21}_{-17}$ \\
 150 & 1.77 & $^{+0.70}_{-0.74}$ & $\times 10^{-3}$ & \quad
2.27 & $^{+0.60}_{-0.48}$ & $\times 10^{-3}$ & \quad\quad $^{ -8.0}_{+11}$ &
$\pm  3.3$ & $\pm  6.4$ & $^{+23}_{-18}$ \\
 160 & 1.26 & $^{+0.49}_{-0.51}$ & $\times 10^{-3}$ & \quad
1.62 & $^{+0.46}_{-0.36}$ & $\times 10^{-3}$ & \quad\quad $^{ -7.2}_{+11}$ &
$\pm  3.2$ & $\pm  6.4$ & $^{+25}_{-20}$ \\
 180 & 6.59 & $^{+2.65}_{-2.60}$ & $\times 10^{-4}$ & \quad
8.61 & $^{+2.79}_{-2.10}$ & $\times 10^{-4}$ & \quad\quad $^{ -5.9}_{ +9.7}$ &
$\pm  3.0$ & $\pm  6.4$ & $^{+30}_{-23}$ \\
 200 & 3.58 & $^{+1.53}_{-1.40}$ & $\times 10^{-4}$ & \quad
4.76 & $^{+1.76}_{-1.29}$ & $\times 10^{-4}$ & \quad\quad $^{ -5.0}_{ +8.9}$ &
$\pm  2.8$ & $\pm  6.4$ & $^{+35}_{-26}$ \\
 250 & 8.78 & $^{+4.67}_{-3.63}$ & $\times 10^{-5}$ & \quad
1.24 & $^{+0.63}_{-0.43}$ & $\times 10^{-4}$ & \quad\quad $^{ -3.5}_{ +7.7}$ &
$\pm  2.6$ & $\pm  6.4$ & $^{+49}_{-34}$ \\
 300 & 2.45 & $^{+1.66}_{-1.11}$ & $\times 10^{-5}$ & \quad
3.76 & $^{+2.50}_{-1.56}$ & $\times 10^{-5}$ & \quad\quad $^{ -2.9}_{ +6.9}$ &
$\pm  2.4$ & $\pm  6.4$ & $^{+66}_{-41}$ \\
 400 & 2.41 & $^{+2.55}_{-1.40}$ & $\times 10^{-6}$ & \quad
4.56 & $^{+4.81}_{-2.58}$ & $\times 10^{-6}$ & \quad\quad $^{ -2.7}_{ +5.4}$ &
$\pm  2.2$ & $\pm  6.4$ & $^{+105}_{ -56}$ \\
\hline \hline
\end{tabular}
\end{center}
\end{table}

\begin{table}[tbp]
\begin{center}
\caption{Same as Table~\ref{tab:sigma1}, but for $\sqrt S = 1.96$ TeV.
\label{tab:sigma2}}
\medskip
\begin{tabular}{cr@{ }l@{ }lr@{ }l@{ }lllll} \hline \hline
\multicolumn{1}{c}{$m_h$ (GeV)} & \multicolumn{3}{c}{$\sigma_{\rm LO}$ (pb)} &
\multicolumn{3}{c}{\quad$\sigma_{\rm NLO}$ (pb)} &
\multicolumn{4}{c}{$\delta\sigma_{\rm NLO}(\delta\mu_F, \delta\mu_R, \delta
y_b,$ PDF) (\%)} \\ \hline
  60 & 9.97 & $^{+9.13}_{-7.68}$ & $\times 10^{-2}$ & \quad
1.39 & $^{+0.41}_{-0.54}$ & $\times 10^{-1}$ & \quad\quad $^{-37}_{+28}$ &
$\pm  5.2$ & $\pm  6.4$ & $^{ +5.5}_{ -5.9}$ \\
  70 & 6.00 & $^{+4.51}_{-4.13}$ & $\times 10^{-2}$ & \quad
8.13 & $^{+2.07}_{-2.47}$ & $\times 10^{-2}$ & \quad\quad $^{-29}_{+23}$ &
$\pm  4.8$ & $\pm  6.4$ & $^{ +6.5}_{ -6.7}$ \\
  80 & 3.71 & $^{+2.38}_{-2.32}$ & $\times 10^{-2}$ & \quad
4.92 & $^{+1.13}_{-1.24}$ & $\times 10^{-2}$ & \quad\quad $^{-23}_{+20}$ &
$\pm  4.5$ & $\pm  6.4$ & $^{ +7.7}_{ -7.7}$ \\
  90 & 2.35 & $^{+1.33}_{-1.35}$ & $\times 10^{-2}$ & \quad
3.08 & $^{+0.66}_{-0.68}$ & $\times 10^{-2}$ & \quad\quad $^{-19}_{+18}$ &
$\pm  4.4$ & $\pm  6.4$ & $^{ +9.2}_{ -8.8}$ \\
 100 & 1.53 & $^{+0.78}_{-0.82}$ & $\times 10^{-2}$ & \quad
1.98 & $^{+0.41}_{-0.40}$ & $\times 10^{-2}$ & \quad\quad $^{-16}_{+16}$ &
$\pm  4.1$ & $\pm  6.4$ & $^{+11}_{-10}$ \\
 105 & 1.24 & $^{+0.60}_{-0.64}$ & $\times 10^{-2}$ & \quad
1.61 & $^{+0.32}_{-0.32}$ & $\times 10^{-2}$ & \quad\quad $^{-15}_{+15}$ &
$\pm  4.1$ & $\pm  6.4$ & $^{+11}_{-11}$ \\
 110 & 1.02 & $^{+0.47}_{-0.51}$ & $\times 10^{-2}$ & \quad
1.31 & $^{+0.27}_{-0.25}$ & $\times 10^{-2}$ & \quad\quad $^{-14}_{+14}$ &
$\pm  4.0$ & $\pm  6.4$ & $^{+12}_{-11}$ \\
 115 & 8.36 & $^{+3.76}_{-4.05}$ & $\times 10^{-3}$ & \quad
1.08 & $^{+0.22}_{-0.20}$ & $\times 10^{-2}$ & \quad\quad $^{-13}_{+14}$ &
$\pm  3.9$ & $\pm  6.4$ & $^{+13}_{-12}$ \\
 120 & 6.90 & $^{+3.00}_{-3.25}$ & $\times 10^{-3}$ & \quad
8.87 & $^{+1.84}_{-1.65}$ & $\times 10^{-3}$ & \quad\quad $^{-12}_{+13}$ &
$\pm  3.8$ & $\pm  6.4$ & $^{+14}_{-12}$ \\
 125 & 5.72 & $^{+2.42}_{-2.63}$ & $\times 10^{-3}$ & \quad
7.34 & $^{+1.55}_{-1.36}$ & $\times 10^{-3}$ & \quad\quad $^{-11}_{+13}$ &
$\pm  3.7$ & $\pm  6.4$ & $^{+15}_{-13}$ \\
 130 & 4.76 & $^{+1.96}_{-2.14}$ & $\times 10^{-3}$ & \quad
6.11 & $^{+1.32}_{-1.14}$ & $\times 10^{-3}$ & \quad\quad $^{-11}_{+12}$ &
$\pm  3.7$ & $\pm  6.4$ & $^{+16}_{-14}$ \\
 140 & 3.34 & $^{+1.32}_{-1.43}$ & $\times 10^{-3}$ & \quad
4.28 & $^{+0.97}_{-0.81}$ & $\times 10^{-3}$ & \quad\quad $^{ -9.3}_{+12}$ &
$\pm  3.5$ & $\pm  6.4$ & $^{+18}_{-15}$ \\
 150 & 2.37 & $^{+0.91}_{-0.98}$ & $\times 10^{-3}$ & \quad
3.04 & $^{+0.73}_{-0.59}$ & $\times 10^{-3}$ & \quad\quad $^{ -8.3}_{+11}$ &
$\pm  3.4$ & $\pm  6.4$ & $^{+20}_{-16}$ \\
 160 & 1.70 & $^{+0.65}_{-0.69}$ & $\times 10^{-3}$ & \quad
2.19 & $^{+0.56}_{-0.45}$ & $\times 10^{-3}$ & \quad\quad $^{ -7.5}_{+10}$ &
$\pm  3.3$ & $\pm  6.4$ & $^{+22}_{-18}$ \\
 180 & 9.10 & $^{+3.47}_{-3.52}$ & $\times 10^{-4}$ & \quad
1.18 & $^{+0.34}_{-0.26}$ & $\times 10^{-3}$ & \quad\quad $^{ -6.2}_{ +9.6}$ &
$\pm  3.1$ & $\pm  6.4$ & $^{+26}_{-20}$ \\
 200 & 5.05 & $^{+2.00}_{-1.92}$ & $\times 10^{-4}$ & \quad
6.66 & $^{+2.19}_{-1.64}$ & $\times 10^{-4}$ & \quad\quad $^{ -5.2}_{ +8.9}$ &
$\pm  2.9$ & $\pm  6.4$ & $^{+31}_{-23}$ \\
 250 & 1.30 & $^{+0.62}_{-0.51}$ & $\times 10^{-4}$ & \quad
1.81 & $^{+0.81}_{-0.57}$ & $\times 10^{-4}$ & \quad\quad $^{ -3.6}_{ +7.6}$ &
$\pm  2.7$ & $\pm  6.4$ & $^{+43}_{-30}$ \\
 300 & 3.83 & $^{+2.30}_{-1.64}$ & $\times 10^{-5}$ & \quad
5.69 & $^{+3.31}_{-2.17}$ & $\times 10^{-5}$ & \quad\quad $^{ -2.9}_{ +6.7}$ &
$\pm  2.5$ & $\pm  6.4$ & $^{+57}_{-37}$ \\
 400 & 4.21 & $^{+3.88}_{-2.26}$ & $\times 10^{-6}$ & \quad
7.48 & $^{+6.85}_{-3.87}$ & $\times 10^{-6}$ & \quad\quad $^{ -2.6}_{ +5.2}$ &
$\pm  2.2$ & $\pm  6.4$ & $^{+91}_{-51}$ \\
 500 & 5.62 & $^{+7.59}_{-3.69}$ & $\times 10^{-7}$ & \quad
1.23 & $^{+1.66}_{-0.80}$ & $\times 10^{-6}$ & \quad\quad $^{ -2.4}_{ +4.6}$ &
$\pm  2.1$ & $\pm  6.4$ & $^{+134}_{ -64}$ \\
\hline \hline
\end{tabular}
\end{center}
\end{table}

\begin{table}[tbp]
\begin{center}
\caption{Same as Table~\ref{tab:sigma1}, but for the LHC ($\sqrt S=14$ TeV
$pp$).\label{tab:sigma3}}
\medskip
\begin{tabular}{cr@{ }l@{ }lr@{ }l@{ }lllll} \hline \hline
\multicolumn{1}{c}{$m_h$ (GeV)} & \multicolumn{3}{c}{$\sigma_{\rm LO}$ (pb)} &
\multicolumn{3}{c}{\quad$\sigma_{\rm NLO}$ (pb)} &
\multicolumn{4}{c}{$\delta\sigma_{\rm NLO}(\delta\mu_F, \delta\mu_R, \delta
y_b,$ PDF) (\%)} \\ \hline
  60 & 2.87 & $^{+3.44}_{-2.32}$ & $\times 10^{0}$ & \quad
4.69 & $^{+1.70}_{-2.35}$ & $\times 10^{0}$ & \quad\quad $^{-49}_{+34}$ &
$\pm  6.8$ & $\pm  6.4$ & $^{ +5.9}_{ -7.3}$ \\
  70 & 2.04 & $^{+2.05}_{-1.51}$ & $\times 10^{0}$ &
3.23 & $^{+0.99}_{-1.35}$ & $\times 10^{0}$ & \quad\quad $^{-40}_{+29}$ &
$\pm  6.3$ & $\pm  6.4$ & $^{ +5.5}_{ -7.0}$ \\
  80 & 1.48 & $^{+1.29}_{-1.01}$ & $\times 10^{0}$ &
2.29 & $^{+0.61}_{-0.82}$ & $\times 10^{0}$ & \quad\quad $^{-34}_{+25}$ &
$\pm  5.8$ & $\pm  6.4$ & $^{ +5.2}_{ -6.7}$ \\
  90 & 1.09 & $^{+0.85}_{-0.70}$ & $\times 10^{0}$ &
1.65 & $^{+0.39}_{-0.52}$ & $\times 10^{0}$ & \quad\quad $^{-30}_{+22}$ &
$\pm  5.6$ & $\pm  6.4$ & $^{ +5.0}_{ -6.4}$ \\
 100 & 8.20 & $^{+5.79}_{-4.91}$ & $\times 10^{-1}$ &
1.22 & $^{+0.26}_{-0.34}$ & $\times 10^{0}$ & \quad\quad $^{-26}_{+19}$ &
$\pm  5.4$ & $\pm  6.4$ & $^{ +4.7}_{ -6.2}$ \\
 105 & 7.15 & $^{+4.83}_{-4.16}$ & $\times 10^{-1}$ &
1.06 & $^{+0.22}_{-0.28}$ & $\times 10^{0}$ & \quad\quad $^{-25}_{+18}$ &
$\pm  5.3$ & $\pm  6.4$ & $^{ +4.6}_{ -6.1}$ \\
 110 & 6.26 & $^{+4.06}_{-3.54}$ & $\times 10^{-1}$ &
9.19 & $^{+1.83}_{-2.34}$ & $\times 10^{-1}$ & \quad\quad $^{-23}_{+18}$ &
$\pm  5.2$ & $\pm  6.4$ & $^{ +4.6}_{ -6.0}$ \\
 115 & 5.50 & $^{+3.44}_{-3.03}$ & $\times 10^{-1}$ &
8.04 & $^{+1.55}_{-1.96}$ & $\times 10^{-1}$ & \quad\quad $^{-22}_{+17}$ &
$\pm  5.1$ & $\pm  6.4$ & $^{ +4.5}_{ -5.9}$ \\
 120 & 4.85 & $^{+2.92}_{-2.61}$ & $\times 10^{-1}$ &
7.05 & $^{+1.31}_{-1.64}$ & $\times 10^{-1}$ & \quad\quad $^{-21}_{+16}$ &
$\pm  5.0$ & $\pm  6.4$ & $^{ +4.4}_{ -5.8}$ \\
 125 & 4.29 & $^{+2.50}_{-2.25}$ & $\times 10^{-1}$ &
6.21 & $^{+1.12}_{-1.39}$ & $\times 10^{-1}$ & \quad\quad $^{-20}_{+16}$ &
$\pm  4.9$ & $\pm  6.4$ & $^{ +4.3}_{ -5.7}$ \\
 130 & 3.81 & $^{+2.15}_{-1.96}$ & $\times 10^{-1}$ &
5.48 & $^{+0.96}_{-1.18}$ & $\times 10^{-1}$ & \quad\quad $^{-19}_{+15}$ &
$\pm  4.9$ & $\pm  6.4$ & $^{ +4.3}_{ -5.6}$ \\
 140 & 3.03 & $^{+1.61}_{-1.49}$ & $\times 10^{-1}$ &
4.32 & $^{+0.71}_{-0.87}$ & $\times 10^{-1}$ & \quad\quad $^{-18}_{+14}$ &
$\pm  4.7$ & $\pm  6.4$ & $^{ +4.1}_{ -5.5}$ \\
 150 & 2.44 & $^{+1.22}_{-1.15}$ & $\times 10^{-1}$ &
3.45 & $^{+0.54}_{-0.65}$ & $\times 10^{-1}$ & \quad\quad $^{-16}_{+13}$ &
$\pm  4.6$ & $\pm  6.4$ & $^{ +4.0}_{ -5.3}$ \\
 160 & 1.98 & $^{+0.94}_{-0.90}$ & $\times 10^{-1}$ &
2.78 & $^{+0.42}_{-0.49}$ & $\times 10^{-1}$ & \quad\quad $^{-15}_{+12}$ &
$\pm  4.5$ & $\pm  6.4$ & $^{ +3.9}_{ -5.2}$ \\
 180 & 1.34 & $^{+0.58}_{-0.57}$ & $\times 10^{-1}$ &
1.86 & $^{+0.26}_{-0.30}$ & $\times 10^{-1}$ & \quad\quad $^{-13}_{+11}$ &
$\pm  4.2$ & $\pm  6.4$ & $^{ +3.8}_{ -4.9}$ \\
 200 & 9.31 & $^{+3.77}_{-3.73}$ & $\times 10^{-2}$ &
1.29 & $^{+0.17}_{-0.19}$ & $\times 10^{-1}$ & \quad\quad $^{-12}_{+10}$ &
$\pm  4.1$ & $\pm  6.4$ & $^{ +3.7}_{ -4.7}$ \\
 250 & 4.19 & $^{+1.45}_{-1.48}$ & $\times 10^{-2}$ &
5.69 & $^{+0.66}_{-0.70}$ & $\times 10^{-2}$ & \quad\quad $^{ -8.9}_{ +8.3}$ &
$\pm  3.8$ & $\pm  6.4$ & $^{ +3.7}_{ -4.4}$ \\
 300 & 2.11 & $^{+0.64}_{-0.67}$ & $\times 10^{-2}$ &
2.83 & $^{+0.31}_{-0.31}$ & $\times 10^{-2}$ & \quad\quad $^{ -7.1}_{ +7.2}$ &
$\pm  3.6$ & $\pm  6.4$ & $^{ +3.9}_{ -4.3}$ \\
 400 & 6.66 & $^{+1.69}_{-1.81}$ & $\times 10^{-3}$ &
8.81 & $^{+0.90}_{-0.88}$ & $\times 10^{-3}$ & \quad\quad $^{ -4.9}_{ +5.6}$ &
$\pm  3.3$ & $\pm  6.4$ & $^{ +4.9}_{ -5.2}$ \\
 500 & 2.56 & $^{+0.58}_{-0.62}$ & $\times 10^{-3}$ &
3.37 & $^{+0.36}_{-0.34}$ & $\times 10^{-3}$ & \quad\quad $^{ -3.6}_{ +5.2}$ &
$\pm  3.1$ & $\pm  6.4$ & $^{ +6.5}_{ -6.5}$ \\
 600 & 1.13 & $^{+0.24}_{-0.25}$ & $\times 10^{-3}$ &
1.47 & $^{+0.17}_{-0.16}$ & $\times 10^{-3}$ & \quad\quad $^{ -2.8}_{ +4.5}$ &
$\pm  2.9$ & $\pm  6.4$ & $^{ +8.6}_{ -8.0}$ \\
 700 & 5.42 & $^{+1.11}_{-1.16}$ & $\times 10^{-4}$ &
7.12 & $^{+0.95}_{-0.85}$ & $\times 10^{-4}$ & \quad\quad $^{ -2.3}_{ +3.8}$ &
$\pm  2.8$ & $\pm  6.4$ & $^{+11}_{-10}$ \\
 800 & 2.80 & $^{+0.59}_{-0.59}$ & $\times 10^{-4}$ &
3.69 & $^{+0.57}_{-0.49}$ & $\times 10^{-4}$ & \quad\quad $^{ -1.9}_{ +3.5}$ &
$\pm  2.7$ & $\pm  6.4$ & $^{+13}_{-11}$ \\
 900 & 1.53 & $^{+0.34}_{-0.32}$ & $\times 10^{-4}$ &
2.03 & $^{+0.36}_{-0.30}$ & $\times 10^{-4}$ & \quad\quad $^{ -1.5}_{ +3.1}$ &
$\pm  2.6$ & $\pm  6.4$ & $^{+16}_{-13}$ \\
1000 & 8.73 & $^{+2.05}_{-1.86}$ & $\times 10^{-5}$ & 1.17 &
$^{+0.23}_{-0.19}$ & $\times 10^{-4}$ & \quad\quad $^{ -1.2}_{ +2.8}$ &
$\pm  2.5$ & $\pm  6.4$ & $^{+19}_{-15}$ \\
\hline \hline
\end{tabular}
\end{center}
\end{table}

\section{Discussion}\label{sec:discussion}

The first puzzle listed in Section~\ref{sec:intro} is solved by choosing the
factorization scale appropriately.  As we showed in the previous section, for
$\mu_F\approx m_h/4$, the $1/\ln(m_h/m_b)$ correction is small and the
$\alpha_S$ correction is modest. Thus perturbation theory in each expansion
parameter individually is well behaved.  Furthermore, we evaluated the
$1/\ln^2(m_h/m_b)$ correction as well, and found that it is vanishingly small
at the relevant factorization scale, providing further evidence that the
perturbative series in $1/\ln(m_h/m_b)$ is well behaved. This series
terminates at this order, while the series in $\alpha_S$ extends to all orders
\cite{Dicus:1998hs,Stelzer:1997ns}.

The second puzzle is also solved by a consideration of the choice of scales,
both factorization and renormalization.  In
Refs.~\cite{Dicus:1998hs,Balazs:1998sb,Carena:2000yx}, all scales were chosen
to be the Higgs-boson mass, $\mu_F=\mu_R=m_h$.   The NLO cross section at the
Tevatron is nearly a factor of ten greater than that obtained by calculating
$gg\to b\bar bh$ and integrating over the momenta of the final-state particles.
However, this factor is much less at lower scales, mostly because the cross
section for $gg\to b\bar bh$ is very scale dependent, and increases
significantly at lower scales.  In contrast, the NLO cross section for $b\bar
b\to h$ has mild scale dependence, and decreases by only about $25\%$ for
$\mu_F\approx m_h/4$.

We have established that the relevant factorization scale for $b\bar b\to h$ is
$\mu_F\approx m_h/4$. It is likely that the relevant factorization scale for
$gg\to b\bar bh$ is also much less than $m_h$, as well as the renormalization
scale of $\alpha_S$.  We have made no attempt to establish the relevant
renormalization scale for the Yukawa coupling in $b\bar b\to h$, and we have
found that our NLO calculation is insensitive to this scale. However, the size
of the $\alpha_S$ correction is less for smaller renormalization scales, which
suggests that the relevant renormalization scale may be less than $m_h$.  Let
us adopt this {\it ansatz}, although we do not have a rigorous justification
for it, in contrast to our derivation of the factorization scale.

As a specific example, we evaluate the cross section for $gg\to b\bar bh$ with
$\mu_F=\mu_R=m_h$ and $\mu_F=\mu_R=m_h/4$ for $m_h=100$ GeV at the Tevatron,
and compare with our NLO calculation of $b\bar b\to h$.  The results are
listed in Table~\ref{tab:compbbgg}.  The order of magnitude difference between
our NLO calculation and $gg\to b\bar bh$ when the scale is $m_h$ is reduced to
about a factor of two for scales close to $m_h/4$.  A factor of two can easily
be accounted for by the fact that our calculation sums collinear logarithms to
all orders, while $gg\to b\bar bh$ produces only the LO collinear logarithm.
This solves the second puzzle listed in the introduction.

\begin{table}[tbp]
\begin{center}
\caption{Cross sections (fb) for $b\bar b\to h$ at NLO and $gg\to b\bar bh$ at
LO for $m_h=100$ GeV at the Tevatron, for two choices of the common
factorization and renormalization scales. The final column gives the ratio of
the cross sections.  The ratio is nearly an order of magnitude for
$\mu_F=\mu_R=m_h$, but is only about a factor of two for $\mu_F=\mu_R=m_h/4$.
\label{tab:compbbgg}}
\medskip
\begin{tabular}{ccc|c} \hline \hline
Scales & $\sigma(b\bar b\to h)$ & $\sigma(gg\to b\bar bh)$ &
$\sigma(b\bar b\to h)/\sigma(gg\to b\bar bh)$ \\
\hline
$\mu_F=\mu_R=m_h$ & 26.6 fb& 3.1 fb & 8.5 \\
$\mu_F=\mu_R=m_h/4$ & 20.8 fb & 9.2 fb & 2.3 \\
\hline \hline
\end{tabular}
\end{center}
\end{table}

It is the desire to sum collinear logarithms that leads one to use $b\bar b\to
h$ as the LO subprocess for the inclusive production of the Higgs boson. The
corrections are of order $1/\ln(m_h/m_b)$ and $\alpha_S$, and we have seen
that they are modest for the appropriate choice of factorization scale. If one
were to use $gg\to b\bar bh$ as the LO subprocess, the expansion parameter
would be $\alpha_S\ln(m_h/m_b)$, and perturbation theory would be poorly
behaved.  Our calculation gives the most accurate and reliable cross section
for the inclusive production of the Higgs boson because it sums these
collinear logarithms to all orders.

It is suggested in Refs.~\cite{Rainwater:2002hm,Spira:2002rd} that the
calculation of $b\bar b\to h$ may overestimate the inclusive cross section,
due to crude approximations inherent in the kinematics, which give rise to
large bottom-quark mass and phase-space effects.  However, the ACOT formalism
\cite{Aivazis:1993pi,Collins:1998rz,Kramer:2000hn} makes no approximations in
either the kinematics or the $b$ mass; we maintained the effect of the $b$
mass exactly.  We find no evidence for any inconsistency in the ACOT
formalism.  Rather, we find that the LO calculation of $gg\to b\bar bh$,
integrated over the momenta of the final-state particles, underestimates the
inclusive cross section when the factorization and renormalization scales are
chosen to be $m_h$.

\section{Conclusions}\label{sec:conclusions}

We have revisited the next-to-leading-order (NLO) calculation of Higgs-boson
production via bottom-quark fusion and solved the two puzzles associated with
that calculation \cite{Dicus:1998hs,Balazs:1998sb}.  We showed that the
appropriate factorization scale for this process is $\mu_F\approx m_h/4$,
rather than $m_h$, as had been previously assumed.  This greatly improves the
convergence of the perturbation series, which was mediocre for $\mu_F=m_h$.
The resulting cross section has mild factorization-scale dependence, and small
renormalization-scale dependence.  It is the most reliable calculation of the
inclusive cross section for Higgs-boson production in association with $b$
quarks.

To support our arguments, we calculated one of the next-to-next-to-leading
(NNLO) corrections (associated with the diagrams in Fig.~\ref{fig:ggbbh}), and
showed that it is vanishingly small for $\mu_F\approx m_h/4$.  The ingredients
exist to calculate the full NNLO cross section, using the results from
Refs.~\cite{Campbell:2002zm,Melnikov:1995yp}. This should yield a cross
section with small factorization- and renormalization-scale dependence.  It
will also provide an additional check of our choice of factorization scale.

The other puzzle we solved also involves the choice of scales, both
factorization and renormalization.  The inclusive cross section for Higgs-boson
production in association with bottom quarks may be approximated by $gg\to
b\bar bh$, integrated over the momenta of the final-state particles. This
yields a result an order of magnitude less than the NLO calculation of $b\bar
b\to h$.  However, this result is very scale dependent, since it is based on a
leading-order calculation.  Choosing scales of order $m_h/4$ rather than
$m_h$, we find that the cross section is comparable to that of $b\bar b\to h$
(see Table~\ref{tab:compbbgg}).  The NLO calculation of $gg\to b\bar bh$ might
support these observations. However, that calculation is not as accurate as one
based on $b\bar b\to h$ for $m_h\gg m_b$, since the latter sums collinear
logarithms to all orders in perturbation theory.

Let us review the existing calculations of Higgs-boson production in
association with bottom quarks.  The relevant calculation depends upon
the final state that is desired.  For the inclusive cross section, the
relevant leading-order (LO) subprocess is $b\bar b\to h$
(Fig.~\ref{fig:bbh}).  The NLO cross section was calculated in
Refs.~\cite{Dicus:1998hs,Balazs:1998sb} and updated in this paper. The
cross section for the production of the Higgs boson accompanied by one
high-transverse-momentum ($p_T$) bottom quark is obtained at LO from
the subprocess $bg\to bh$ (Fig.~\ref{fig:gbhb}), which is calculated
at NLO in Ref.~\cite{Campbell:2002zm}.  This process is particularly
promising due to the ability to tag the $b$ quark in the final
state. Finally, the cross section for the production of the Higgs
boson accompanied by two high-$p_T$ $b$ quarks is obtained at LO from
the subprocesses $gg,q\bar q\to b\bar bh$ (Fig.~\ref{fig:ggbbh}). This
process has been calculated only at LO thus far, but the ingredients
exist to provide the NLO cross section
\cite{Beenakker:2001rj,Beenakker:2002nc,Reina:2001sf,Reina:2001bc,Dawson:2002tg}.
Although this process has been the most studied, it is likely that
$bg\to bh$ is the more promising, due to its larger cross section.
The inclusive cross section, $b\bar b\to h$, obtained in this paper,
is useful when the Higgs boson can be identified above backgrounds
without the need to tag $b$ quarks, such as $h\to \tau^+\tau^-,
\mu^+\mu^-$.  It has the advantage of having the largest cross
section, since it is inclusive of the other two processes.

\section*{Acknowledgments}

We are grateful for conversations and correspondence with E.~Braaten,
A.~Connolly, D.~Froidevaux, T.~Junk, P.~Lepage, P.~Mackenzie, and
E.~Richter-Was.  F.~M. warmly thanks the Department of Physics of the ``Terza
Universit\`a di Roma'' for the kind hospitality and support.  This work was
supported in part by the U.~S.~Department of Energy under contracts
Nos.~DE-FG02-91ER40677 and DE-AC02-76CH03000.

\end{document}